\documentclass[aps,prd,twocolumn,reprint,nofootinbib,showkeys,showpacs,shttps,scriptaddress,longbibliography,floatfix]{revtex4}

\usepackage{amsmath}
\usepackage{amssymb}
\usepackage[american]{babel}
\usepackage{booktabs}
\usepackage{dcolumn}  
\usepackage[T1]{fontenc}  
\usepackage{graphicx}  
\usepackage[colorlinks]{hyperref}  
\usepackage[utf8]{inputenc}  
\usepackage{multirow}
\usepackage{txfonts}  
\usepackage{url}
\usepackage{color}
\usepackage[dvipsnames]{xcolor}
\usepackage{float}
\usepackage{lipsum}
\newcolumntype{d}[1]{D{.}{.}{#1}}
\newcolumntype{v}[1]{D{,}{,\ }{#1}}

\graphicspath{{Figuras/}}


\renewcommand {\arraystretch}{1.5}

\hypersetup{
	citecolor = blue,
	linkcolor = RoyalPurple,
	urlcolor = blue
}

\begin{document}	
		
	\title{Strong lensing systems and galaxy cluster observations as probe to the cosmic distance duality relation}

	\author{R. F. L. Holanda}
	\email{holandarfl@fisica.ufrn.br}
	\affiliation{Departamento de F\'{\i}sica, Universidade Federal do Rio Grande do Norte,Natal - Rio Grande do Norte, 59072-970, Brasil}
	\affiliation{Departamento de F\'{\i}sica, Universidade Federal de Campina Grande, 58429-900, Campina Grande - PB, Brasil.}
	\affiliation{Departamento de F\'{\i}sica, Universidade Federal de Sergipe, 49100-000, Aracaju - SE, Brazil}
		
		\author{F. S. Lima}
	\email{felixsilva775@live.com}
	\affiliation{Departamento de F\'{\i}sica, Universidade Federal de Sergipe, 49100-000, Aracaju - SE, Brazil}

	\author{Akshay Rana}
	\email{montirana92@gmail.com}
	\affiliation{St. Stephen's College, University of Delhi, Delhi  110007, India}
	
		\author{Deepak Jain}
	\email{djain@ddu.du.ac.in}
	\affiliation{Deen Dayal Upadhyaya College, University of Delhi, New Delhi 110078, India}

	
	\date{\today}

	\begin{abstract}
	{In this paper, we use  large scale structure observations to test the redshift dependence of cosmic distance duality relation (CDDR), $D_{\rm L}(1+z)^{-2}/D_{\rm A}=\eta(z)$}, with $D_{\rm L}$ and $D_{\rm A}$, being the luminosity and angular diameter distances, respectively. In order to perform the test, the following data set are considered: strong lensing systems and galaxy cluster measurements (gas mass fractions). No specific cosmological model is adopted, only a flat universe is assumed.  { By considering two $\eta(z)$ parametrizations, It is observed that  the CDDR remain redshift independent within $1.5\sigma$ which is in full agreement with other recent tests involving cosmological data}. It is worth to comment that our results are independent of the baryon budget of galaxy clusters.
	\end{abstract}

	\maketitle
	
	\section{Introduction}\label{sec:intro}
	
 Two types of distance  are mostly used in cosmology : the luminosity distance, $D_{\rm L}$, and the angular diameter distance, $D_{\rm A}$. It is known that  $D_{\rm L}$  is a distance measurement associated with an object based on the decrease of its brightness, and $D_{\rm A}$ is associated with the measurement of the angular size of the object projected on the celestial sphere. Measuring distances in cosmology is of crucial importance  as it  relates observational data with the theoretical assumptions. These cosmological distances are connected by a relation known as the cosmic distance duality relation (CDDR), it is the astronomical version of the  Etherington's reciprocity law and it is written as:  $\frac{D_{\rm L}(z)}{D_{\rm A}(z)(1+z)^{2}} = \eta= 1$ \cite{Etherington1933,Etherington2007}. Actually, the CDDR is obtained in the context of Friedmann-Lemaître-Robertson-Walker metric which holds for general metric theories of gravity in any background \cite{Bassett2004,Ellis2009}. A little deviation from $\eta=1$ may indicate the possibility of a new physics \cite{Avgoustidis2010,Jaeckel2010,Bassett2004,Avgoustidis2010,Jaeckel2010,Hees2014}. Besides, the presence of systematic errors in observations also can violate the validity of CDDR  \cite{Bassett2004,Avgoustidis2010,Holanda2013}.
	
	In the last decade,  different methods have been proposed to test the validity of the CDDR due to  the improvement in the quantity and the quality of astronomical data. These methods can be divided in two classes: cosmological model-dependent tests based on the $\Lambda$ cold dark matter ($\Lambda$CDM) framework (see the Refs. \cite{DEBERNARDIS2006,Uzan2004,Avgoustidis2010,Holanda2011,Piazza2016}) and cosmological model-independent. Several astronomical data sets  have been used to probe the CDDR, for instance: angular diameter distance of galaxy clusters, galaxy cluster gas mass fraction, type Ia supernovae (SNe Ia), strong gravitational lensing (SGL), cosmic microwave background, gamma ray bursts, radio compact sources, cosmic microwave background radiation, baryon acoustic oscillations, gravitational waves, etc.  \cite{Holanda2010,Lima2011,Li2011,gon,Meng2012,Holanda2012,Yang2013,Liang2013,Shafieloo2013,Zhang:2014eux,SantosdaCosta2015,Jhingan2014,Chen2015,Holanda2016,Rana2016,Liao2016,Holanda2016b,Holanda2017,Rana2017,Lin2018,Fu2019,Ruan2018,Holanda2019,Chen2020,Zheng2020,Kumar:2020ole,Hu2018}. In particular, the Ref. \cite{Holanda2012} used only massive galaxy clusters observations (Sunyaev-Zeldovich effect and X-ray surface brightness observations)  to test the CDDR.
	
	In order to test { the redshift dependence of CDDR,} the basic approach has been to consider a  modified expression, given by $\frac{D_{\rm L}(z)}{D_{\rm A}(z)(1+z)^{2}} = \eta(z)$, and to obtain constraints on  $\eta(z)$ functions. In the Ref.\cite{HolandaWilliam} a Bayesian model comparison of various $ \eta(z) $ functions  are used, such as: $\eta=\eta_0$, $\eta(z)=1+ \eta_0 z$, $\eta(z)=1+ \eta_0z/(1+z)$, $\eta(z)=\eta_0 + \eta_1z$ and  $\eta(z)=\eta_0 + \eta_1z/(1+z)$. The idea was to estimate the Bayesian evidence and compute the Bayes factor for each $\eta(z)$ function with respect to $\eta=\eta_0$. The results favor  the $\eta(z) = constant$  as the standard model. On the other hand, the Ref. \cite{SantosdaCosta2015} applied a non-parametric method, namely, Gaussian process, to test the CDDR based on galaxy clusters observations and $H(z)$ measurements (see also the Ref. \cite{Chen2015}). The CDDR validity has been verified, at least, within $2\sigma$ c.l. in  the last decade . However, it is  worth  to stress that the current analysis cannot distinguish which $\eta(z)$ function better describes the data (see \cite{HolandaWilliam}) for details.

	{ In this paper, we show that strong lensing systems and galaxy cluster observations (gas mass fraction) can be used to study the redshift dependence of the CDDR  and such constraints are competitive with those ones obtained from other  cosmological observations}. The data sets used  are: SGL subsample from the original data set compiled by the Ref. \cite{Leaf2018lfu}. The galaxy clusters observations correspond to 40 gas mass fraction from the Ref. \cite{Mantz2014}.  The $\eta(z) $ functions used in this paper are: $\eta_I(z)=1+ \eta_0 z$, $\eta_{II}(z)=1+ \eta_0z/(1+z)$.  
	
	This paper is organised as follows: In Sec.(II), we describe the data used in this work, SGL systems and galaxy clusters observations. The methodology is discussed in the Sec.(III). In the Sec.(IV), the $\eta(z)$ functions are presented. The main results of the statistical analysis are highlighted  in the Sec.(V) and in the Sec.(VI),  we describe the main conclusions of this work.

\section{Cosmological data}

\begin{itemize}
\item We use the most recent x-ray mass fraction measurements of $40$ galaxy clusters in redshift range $0.078 \leq z \leq 1.063$  \cite{Mantz2014}. These authors measured the gas mass fraction in spherical shells at radii near $r_{2500}$, rather than integrated at all radii (less than $r_{2500}$). Hence the theoretical uncertainty in the gas depletion obtained from hydrodynamic simulations is reduced \cite{Mantz2014,Planelles2013}.  Moreover, the bias in the mass measurements from X-ray data arising by assuming hydrostatic equilibrium was calibrated by robust mass estimates for the target clusters (see also \cite{app}). 
\item We also consider a subsample (101 points) from a specific catalog containing 158 confirmed sources of strong gravitational lensing \cite{Leaf2018lfu}. This complete compilation includes 118 SGL systems identical to the compilation of \cite{Cao2015} which are obtained from SLOAN Lens ACS, BOSS Emission-line Lens Survey (BELLS) and Strong Legacy Survey SL2S along with 40 new systems recently discovered by SLACS and pre-selected by \cite{Shu2017} (see Table I in \cite{Leaf2018lfu}). The 101 points used here are from the original sample (158) whose redshifts are lower than $z=1.063$ and with the quantity $D$, distance ratio (see next section), compatible with $D=1$ within 1$\sigma$ c.l. ($D > 1$ is a no physical region). 
\end{itemize}

\section{Methodology}

\subsection{Strong Gravitational Lensing Systems}

As it is largely known, strong gravitational lensing systems, one of the predictions of GR \cite{lentes},  is a purely gravitational phenomenon occurring when the source ($s$), lens ($l$), and observer ($o$) are at the same signal line forming a structured ring called the Einstein radius ($\theta_E$) \cite{lentes}. Usually,  a lens can be a foreground galaxy or cluster of galaxies positioned between a source--Quasar, where the multiple-image separation from the source only depends on the lens and source angular diameter distance. It is important to note that such systems have recently become a powerful astrophysical tool. For instance, the Hubble constant has been estimated by time-delay measurements \cite{Refsdal,Treu,Wong,Liao,Liao2,kocha}, the CDDR was tested by using these systems in the Refs. \cite{Liao2016,Holanda2017b,Ruan2018,rana}. Moreover, properties of SGL also can restrict the deceleration parameter of the universe \cite{gott}, space-time curvature \cite{jzqi,rana}, the cosmological constant \cite{fuku}, the speed of light \cite{luz}, and many more. 

However, it is important to stress that  the constraint obtained from SGL  depends on a model of mass distribution of lens. In the most simple assumption, the singular isothermal sphere (SIS) model, the Einstein radius $\theta_E$ is given by \cite{lentes}

\begin{equation}
\theta_E = 4\pi \frac{D_{A_{ls}}}{D_{A_{s}}} \frac{\sigma_{SIS}^{2}}{c^2},
\end{equation}
where $D_{A_{ls}}$ is the angular diameter distance of the lens to the source, $D_{A_{s}}$ the angular diameter distance of the observer to the source, $c$ the speed of light, and $\sigma_{SIS}$ the velocity dispersion caused by the lens mass distribution. It is important to note here that $\sigma_{SIS}$ is not exactly equal to the observed stellar velocity dispersion ($\sigma_0$) due to a strong indication, via X-ray observations, that dark matter halos are dynamically hotter than luminous stars. 

A method developed by the Ref. \cite{Liao2016} provided a powerful test for the CDDR using SGL systems and SNe Ia. The method is based on equation (3) for lenses and an observational quantity defined by

\begin{equation}
D \equiv \frac{D_{A_{ls}}}{D_{A_s}} = \frac{ \theta_E c^2}{4\pi \sigma_{SIS}^{2}}.
\end{equation}
By assuming a flat universe with the comoving distance between the lens and the observer is given as $r_{ls} = r_s-r_l$ \cite{bartel}, and using the relations $r_s = (1 + z_s) D_{A_s}$, $r_l = (1 + z_l) D_{A_l}$, $r_{ls} = (1 + z_s) D_{A_{ls}}$, one can obtain

\begin{equation}
D = 1-\frac{(1+z_l)}{(1+z_s)}\frac{D_{A_{l}}(z)}{D_{A_{s}}(z)}.
\end{equation}
By using a deformed CDDR, the above equation  may be written as: 
\begin{equation}
(1-D)\frac{D_{L_{s}}(z)}{D_{L_{l}}(z)} =\frac{(1+z_s)\eta_s(z)}{(1+z_l)\eta_l(z)}. 
\end{equation}

	\begin{figure*}[t]
		\centering
		\includegraphics[scale=0.63]{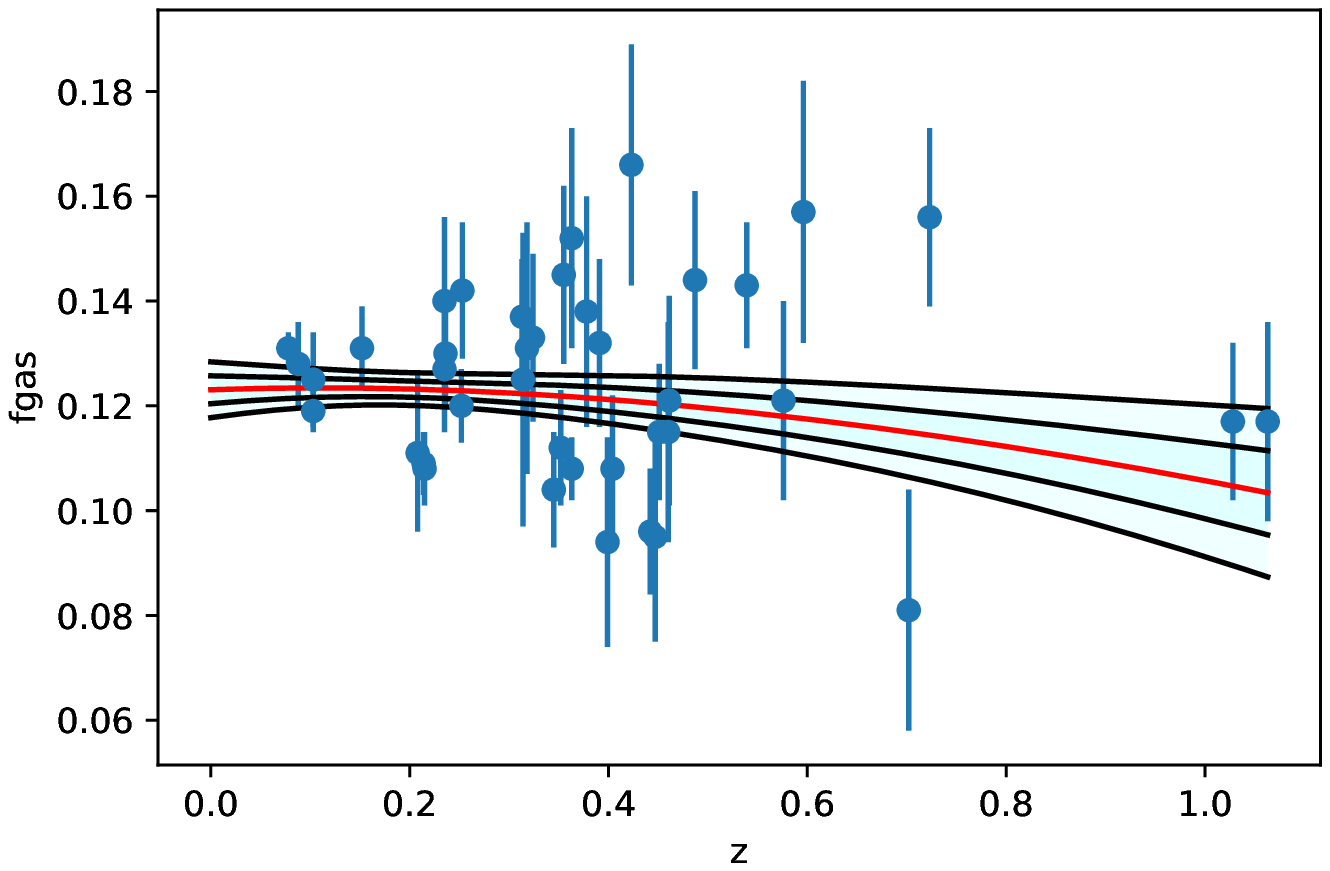}
		\includegraphics[scale=0.30]{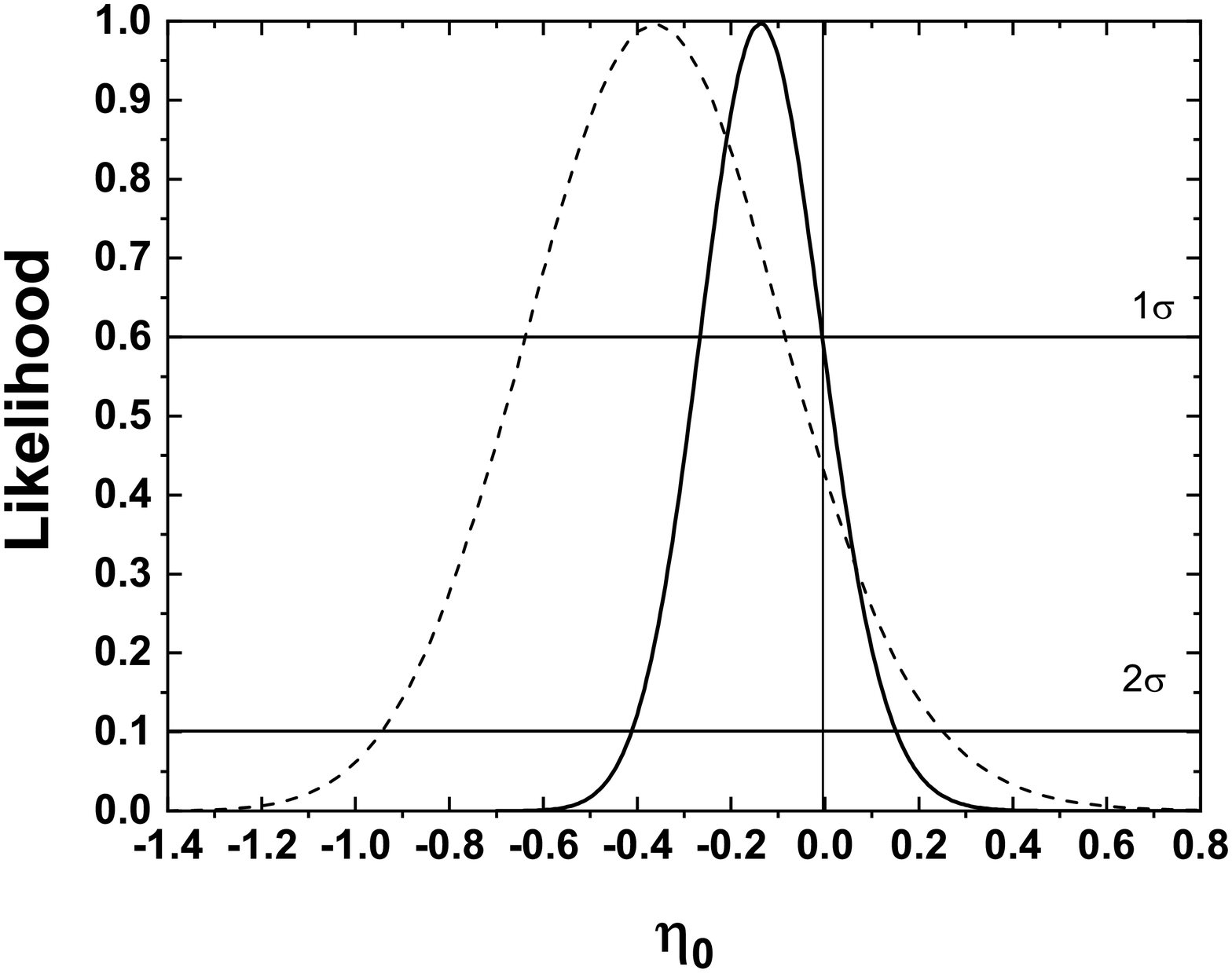}
		\caption{\label{fig:regions2}{\it Left - }Smooth plot of fgas versus redshift obtained  after applying Gaussian process.The red line shows the best fit line while dark and light regions are the 1$\sigma$ and 2$\sigma$ confidence bands respectively.  {\it Right - }The posteriors distributions for the $\eta(z)$ functions. The solid and black lines correspond to Eqs.(9) and (10).}
	\end{figure*}

\subsection{Gas mass fraction}

Briefly, the cosmic gas mass fraction is defined as $f_{gas}=\Omega_b/\Omega_M$. The assumed constancy of this quantity within massive, relaxed clusters can be used to constrain cosmological parameters by using the following expression \cite{Allen2008,gon} 
\begin{equation}
\label{GasFrac}
f_{gas}(z) = N \left[\frac{D_L^{*}(z){D_{A}^{*1/2}(z)}}{D_L(z){D_{A}^{1/2}(z)}}\right]\;,
\end{equation}
{where the observations are taken in X-ray band, the normalization factor $N$ carries all the information about the matter content in the cluster, such as stellar mass fraction,  non-thermal pressure and the depletion parameter $\gamma$, which indicates the amount of cosmic baryons that are thermalized within the cluster potential (this term will be detailed in next section).} The asterisk denotes the corresponding quantities in the fiducial model used in the observations to obtain the $f_{gas}$ (a flat $\Lambda$CDM model with Hubble constant $H_0=70$ km s$^{-1}$ Mpc$^{-1}$ and the present-day matter density parameter $\Omega_M=0.3$).  It is important to comment that the ratio in the brackets of Eq.(5) computes the expected variation in $ f_{gas}^\mathrm{ref}\left(z\right)$ when the underlying cosmology is varied. This term also accounts for deviations in the geometry of the Universe from the reference model, which makes the analysis model-independent (see \cite{Allen2008}  for more details when $\eta = 1$ is assumed in the reference model).


 Our method is completely based on the recent results from the Refs.\cite{gon,Holanda2012}. In Ref.\cite{gon}, the authors showed that the gas mass fraction measurements  extracted from X-ray data are affected by a possible departure of $\eta =1$ and the Eq.(5) must be rewritten as
\begin{eqnarray}
\label{GasFrac3}
f_{gas}(z) = N \left[\frac{\eta(z)^{1/2}D_L^{*3/2}(z)}{D_{L}^{3/2}(z)}\right].
\end{eqnarray}
 The $\eta$ parameter appears after using the deformed  CDDR relation in  the denominator. So, the luminosity distance in galaxy cluster redshift can be calculated by:
 
 \begin{equation}
D_L(z) = N^{2/3} \left[\frac{\eta(z)^{1/3}D_L^{*}}{f_{gas}^{2/3}(z)}\right].
\end{equation}
Finally, by using equations (7) and (4), one may obtain the key equation for our test :
\begin{equation}
(1-D) = \frac{(1+z_s)}{(1+z_l)}\frac{\eta_{s}^{2/3}(z)}{\eta_{l}^{2/3}(z)}\frac{D^*_{L_{l}}(z)f_{gas}^{2/3}(z_s)}{D^*_{L_{s}}(z)f_{gas}^{2/3}(z_{l})}.
\end{equation}
or
\begin{equation}
\frac{\eta^{2/3}_s(z)}{\eta^{2/3}_l(z)} = (1-D)\frac{(1+z_l)}{(1+z_s)}\frac{D^*_{L_{s}}(z)f_{gas}^{2/3}(z_l)}{D^*_{L_{l}}(z)f_{gas}^{2/3}(z_{s})}.
\end{equation}

Our method is independent on $N$ if it is a constant (see next section for details). { The asterisk denotes the corresponding quantities calculated in the fiducial model used in the observations to obtain the $f_{gas}$ (a flat $\Lambda$CDM model with Hubble constant $H_0=70$ km s$^{-1}$ Mpc$^{-1}$ and the present-day matter density parameter $\Omega_M=0.3$).}

As one may see, in order to put limits on $\eta_0$ it is necessary to have  gas mass fraction measurements  on lens and source redshifts for each SGL system. {{This is obtained by applying a model independent non-parametric smoothing technique, Gaussian Process, (see \cite{Holsclaw,Shafieloo,seikel}) on the 40 gas mass fraction obtained by the Ref.\cite{Mantz2014} (see fig.1-left). Gaussian Process is a widely used smoothing method in which the complicated parametric relationship is replaced by parametrizing a probability model over the data. In mathematical terms, it is a distribution over functions, characterized by a mean function and covariance function. In order to avoid model dependence appearing through the choice of  prior mean function, we have chosen it to be zero. It is quite common choice of mean function as one can always normalize the data so it has zero mean. This method comes with a few inherent underlying assumptions as: each observation is an outcome of an independent Gaussian distribution belonging to the same population and the outcome of observations at any two redshifts are correlated due to their nearness to each other. The correlation function between two redshifts (say $z$ \& $z'$ ) used in this analysis is  square exponential kernel, given by,}}

\begin{equation}
    k(z,z')= \sigma_f^2 \exp{-\dfrac{(|z-z'|)^2}{2l^2}}
\end{equation}
{Here, $\sigma_f$ and $l$ are two hyperparameters which are calculated by maximizing the corresponding marginal log-likelihood probability function of the distribution. }

	\section{Parametrizations}\label{sec:parameters}
	 The $\eta(z)$ parametrizations considered here are  \cite{Holanda2010,Holanda2011,Liang2013,Li2011}:

	\begin{equation}\label{eq:M2}
	\eta_I(z) = 1 + \eta_{0}z, 
	\end{equation}
	
	\begin{equation}\label{eq:M3}
	\eta_{II}(z) = 1 + \eta_{0}\frac{z}{1+z}, 
	\end{equation}

	These are the main $\eta(z)$ functions widely used in the literature as it effectively parametrize our ignorance of the underlying process responsible for a possible CDDR violation.   
	
	\begin{table*}[ht]
\caption{A summary of the current constraints on the $\eta_0$ and $\epsilon$ parameters from different observables.}
\label{tables1}
\par
\begin{center}
\begin{tabular}{|c||c|c|c|c|}
\hline\hline  Reference &  Data Sample &$1+\eta_0z$ & $1+\eta_0z/(1+z)$& $(1+z)^{\epsilon}$  
\\ \hline\hline 
\cite{Avgoustidis2012}*\footnote{The symbol ``*'' means 2$\sigma$ error bars } & SNe Ia + $\Lambda$CDM + $H(z)$&-& - &$ -0.04^{+0.08}_{-0.07}$ \\ 
\cite{Nair2012} &BAO + SNe Ia &  $-0.098 \pm 0.084$    & $-0.151 \pm 0.155$ & - \\
\cite{gon}* & Gas mass fractions + SNe Ia  &$-0.03^{+1.03}_{-0.65}$& $-0.08^{+2.28}_{-1.22}$ & - \\
\cite{Holanda2012}* & Only Gas mass fractions & $-0.06 \pm 0.16$ & $-0.07 \pm 0.24$ &- \\
\cite{Yang2013}& ADD   + SNe Ia      & $0.16^{+0.56}_{-0.39}$    & - & - \\
\cite{Holanda2013} & SNe Ia + $H(z)$ &-&-& $0.017 \pm 0.055$  \\
\cite{Jhingan2014} & Radio galaxies + SNe Ia &  $-0.180 \pm 0.244$ & $-0.415 \pm 0.632$ & -  \\
\cite{SantosdaCosta2015}* & ADD  + $H(z)$   & $-0.100^{+0.117}_{-0.126}$ & $-0.157^{+0.179}_{-0.192}$& -  \\
\cite{SantosdaCosta2015}* & Gas mass fraction  + $H(z)$ & $0.062^{+0.168}_{-0.146}$ & $-0.166^{+0.337}_{-0.278}$ & -  \\
\cite{Chen2015} & ADD + SNe Ia + $H(z)$ &$0.07 \pm 0.08$ & $0.15 \pm 0.18$ &-  \\
\cite{Puxon2015} & SNe Ia + BAO & $-0.027 \pm 0.064$ & $-0.039 \pm 0.099$& - \\
\cite{Liao2016} &SGL + SNe Ia & $ -0.005^{+0.351}_{-0.215}$ & $ - $ & - \\
\cite{Holanda2016b}\footnote{Planck results} & SGL (SIS) + SNe Ia + $\Lambda$CDM & $0.05 \pm 0.15$ & $0.09 \pm 0.3$& - \\
\cite{Holanda2016b}$^{b}$ & SGL (PLaw) + SNe Ia + $\Lambda$CDM & $0.08 \pm 0.22$ & $0.06 \pm 0.33$&  \\
\cite{Holanda2016b}\footnote{WMAP9 results} & SGL (SIS) + SNe Ia + $\omega(z)$CDM & $0.01 \pm 0.22$ & $0.017 \pm 0.28$& -\\
\cite{Holanda2016b}$^{c}$ & SGL (PLaw) + SNe Ia + $\omega(z)$CDM & $0.054 \pm 0.29$ & $0.0035 \pm 0.3$& - \\
\cite{Fu2017}\footnote{Crossing Statistic with Smoothing method} & SGL (SIS) + SNe Ia + GRB & $-0.072\pm 0.023$ & $-0.173\pm 0.037$ & - \\
\cite{Fu2017}$^d$ & SGL (PLaw) + SNe Ia + GRB & $0.025^{+0.025}_{-0.024}$ & $0.065^{+0.075}_{-0.066}$ & - \\
\cite{Holanda2017b} & SGL (PLaw) + SNe Ia + GRB & $0.00 \pm 0.10$ & $-0.036^{+0.37}_{-0.32}$& $-0.16^{+0.24}_{-0.52}$ \\
\cite{Holanda2017b} & SGL (SIS) + SNe Ia + GRB & $0.15 \pm 0.13$ & $-0.18^{+0.45}_{-0.65}$& $0.27^{+0.22}_{-0.38}$ \\
\cite{Ruan2018} & SGL (SIS) + HII-GP & $0.0147^{+0.056}_{-0.066}$ & - & - \\
\cite{Li2018}\footnote{Markov chain Monte Carlo methods} & SNe Ia + RS& $-0.06\pm 0.05$ & $-0.18\pm 0.16$  & - \\
This paper\footnote{Considering 101 SGL systems with $D\pm \sigma_D < 1$.} & SGL + galaxy clusters & $-0.13 \pm 0.13$&  $-0.36 \pm 0.34$ & -\\
This paper\footnote{Considering 87 SGL systems with $D < 1$.} & SGL + galaxy clusters & $-0.01 \pm 0.18$&  $-0.13 \pm 0.24$ & -\\

\hline\hline
\end{tabular}
\end{center}
\end{table*}

	\section{Analysis and results}\label{sec:beyesian}

Since recent several studies have shown that slope of density profiles of individual galaxies show a non-negligible scatter from the SIS model (see the Ref. \cite{elilensing} and references there in),  hence a  more  general approach to describe the lensing systems is considered: one with spherically symmetric mass distribution in lensing galaxies in favor of power-law index $\Upsilon$, $\rho \propto r^{-\Upsilon}$ (PLaw). Under this assumption, the equation (2) is written as \cite{lentes}:

\begin{equation}
D = \frac{\theta_E c^2}{4 \pi \sigma_{ap}^{2}}f(\theta_E, \theta_{ap}, \Upsilon),
\end{equation}
where $f (\theta_E, \theta_{ap}, \Upsilon)$ is a  function which depends on the Einstein's radius $\theta_E$, the angular aperture $\theta_{ap}$, used by certain gravitational lensing surveys, and the power-law index $\Upsilon$. If $\Upsilon = 2$,  it gives the SIS model. In this paper, we use a flat prior on the factor $\Upsilon$ ($1.75 \leq \Upsilon \leq 2.2$). Thus, the uncertainty related to Eq.(11) is given by:

\begin{equation}
\sigma_{D} = D\sqrt{4\Bigg(  \frac{\sigma_{\sigma_{ap}}}{\sigma_{ap}}  \Bigg)^2 + (1-\Upsilon)^2 \Bigg(  \frac{\sigma_{\theta_E}}{\theta_E}  \Bigg)^2}.
\end{equation}
Following the approach used by \cite{grillo}, Einstein's radius uncertainty follows $\sigma_{\theta_E} = 0.05 \theta_E$ ($5\%$ for all systems). As mentioned in  ref.\cite{Leaf2018lfu}, it is necessary to add  $12.22\%$ of intrinsic error associated to $D$ measurement. As the random variation in galaxy morphology is almost Gaussian, the authors of Ref. \cite{Leaf2018lfu} found that an additional error term of about $12.22\%$ is necessary to have $68.3\%$ of the observations to lie within $1\sigma$ of the best-fit $\omega$CDM model (see also the Ref. \cite{lentes}). As commented earlier, we have considered in our analyses only the SGL systems whose the quantity $D$ (by fixing $\Upsilon =2$ ) is compatible with unity within 1$\sigma$ c.l.. 	
	
On the other hand, the  term $N$ in Eq.(5) is $N= \gamma(z) K(z) (\Omega_b/\Omega_M)$, where $K(z)$ quantifies inaccuracies in instrument calibration as well as any bias in the mass measured due to substructure, bulk motions and/or non-thermal pressure in the cluster gas.  The $K(z)$ parameter for this sample was estimated to be  $K=0.96 \pm 0.12$ and no significant trend  with mass, redshift or the morphological indicators were verified \cite{app}.  The depletion factor, $\gamma$, was estimated for this sample to be $\gamma = 0.85 \pm 0.05$ (see fig. (6) in the Ref.\cite{Mantz2014}). No evolution for the $\gamma$ parameter has been observed (see also the Refs. \cite{holandafelipe,holandaso}). As commented earlier, if $N$ is constant (it is assumed constant in this paper), our results  are independent of the baryon budget of galaxy clusters.
	
Finally, in order to obtain  constraints on $\eta_0$ for the two functions, we use the Eq.(9) and evaluate the statistical analysis by defining the likelihood distribution function, ${\cal{L}} \propto e^{-\chi^{2}/2}$, where 
\begin{eqnarray}
\chi^2 &=& \sum_{i=1}^{101}\frac{\left(\frac{\eta^{2/3}_s(z)}{\eta^{2/3}_l(z)} - (1-D)\frac{(1+z_l)}{(1+z_s)}\frac{D^*_{L_{s}}(z)f_{gas}^{2/3}(z_l)}{D^*_{L_{l}}(z)f_{gas}^{2/3}(z_{s})} \right)^2}{\sigma_{obs}^2}\nonumber,
\end{eqnarray} 
 where $\sigma_{obs}^2$ correspond to error from the $D$ measurements and cluster gas mass fractions, being obtained from usual method of error propagation. As commented earlier, the asterisk denotes the corresponding quantities in the fiducial model used in the observations to obtain the $f_{gas}$ for each galaxy cluster (a flat $\Lambda$CDM model with Hubble constant $H_0=70$ km s$^{-1}$ Mpc$^{-1}$ and the present-day matter density parameter $\Omega_M=0.3$). { In order to obtain the values of gas mass fraction and its respective 1$\sigma$ error at each $z_s$ and $z_l$, we apply Gaussian Process on gas mass fraction data as shown in the Fig. 1 (left).}  It is important to highlight an in-build assumption about the data i.e; all the direct measurements of $f_{gas}$ mass fraction and strong gravitational lensing are independent and a random outcome of the Gaussian distribution.  We observed the same in most of the published works working with the similar dataset \cite{melia,wang,amante, lizardo,cao}.   As to apply Bayesian analysis, one need to know the parent probability distribution function associated with each observation. If that is not known (which is the case with most of the astronomical observations and measures) then Gaussian is found to be the best choice as it is strongly supported by central limit theorem \cite{wall, licia, sivia}.

In Fig.(1-right) the likelihoods for the two $\eta(z)$ functions are plotted.  For both the   $\eta(z)$ functions , I and II , the results are (at $1\sigma$ and $2\sigma$): $\eta_0 = -0.13 \pm 0.13 \pm 0.27$ and $\eta_0 =-0.36 \pm 0.34 \pm 0.61 $, respectively.{ As one may see,  the CDDR remain redshift independent   within $2\sigma$ c.l.} A summary of the current constraints on the $\eta_0$ and from different cosmological observable are displayed in the Table I (some authors also have used another deformed CDDR, such as the fifth column in the table).

 We also have performed the analyses by considering another SGL subsample (and the corresponding $f_{gas}$ measurements) containing SGL systems whose the $D$ quantity (fixing $\Upsilon=2$) is lower than 1. This new subsample contains 87 SGL systems and the results for $\eta(z)$ functions given by equations (11) and (12) are, respectively: $\eta_0=-0.01 \pm 0.18$  and $\eta_0= -0.13 \pm 0.24$ at 1$\sigma$ c.l. (by marginalizing on $\Upsilon$). As one may see, these results  are also in full agreement with the CDDR validity.

	\section{Conclusions}\label{sec:conclusions}
 
 {In this work, a new method was proposed to test the redshift dependence of  cosmic distance duality relation, $D_L/D_A=(1+z)^2\eta(z)$ }. Galaxy clusters data (gas mass fraction) plus strong gravitational lensing systems are used for that purpose. No specific cosmological model was used in analysis. Unlike most tests involving galaxy cluster data, our results are independent of the baryon budget of a clusters (if it is independent on cluster redshift).{ For both the two $\eta(z)$ parameterizations considered,  the CDDR  remain redshift independent  within $\approx 1.5\sigma$ c.l. (see Table I)}. 

On the other hand, it is worth to comment that by taking
the  Planck best-fitted cosmology, the Ref. \cite{Caob}
considered SGL observations and relaxed the assumption that stellar luminosity and total mass distribution
follows the same power-law. Their results indicate that a model in which mass traces light
 is rejected at $>$ 95\% c.l.. Moreover, the authors also find  the presence of dark matter in the form of
a mass component distributed differently from the light (see also the Ref.\cite{Schwab}). One may check the consequences  of relaxing the rigid assumption that the stellar luminosity and total mass distributions follow the same power law on our approach.

	In the near future, it is expected that several surveys (Erosita, EUCLID mission, Pan-STARRS, LSST, JDEM) will
discover thousands of strong lensing systems and galaxy clusters. So, the method proposed here,  will put 
stringent limits on the  cosmological parameter $\eta_0$. It is important to note that our method considered a flat universe and the role of curvature on our results will be an interesting extension of this work when better data are available.

	\begin{acknowledgements}
	 R. F. L. Holanda thanks financial support from Conselho Nacional de Desenvolvimento Cientifico e Tecnologico (CNPq) (No. 428755/2018-6 and 305930/2017-6).
	\end{acknowledgements}


\begin{thebibliography}{}
\bibitem{Etherington1933}{I.Etherington, The London, Edinburgh, and Dublin Philos. Mag. and J. of Scien. 15, 761 (1933)}
\bibitem{Etherington2007}{I. M. H. Etherington, Gen. Rel. and Grav. 39,
1055 (2007).}
\bibitem{Bassett2004}{B. A. Bassett and M. Kunz, Phys. Rev. D 69, 101305 (2004).}
\bibitem{Ellis2009}{G. F. R. Ellis, Gen. Rel. and Grav. 41, 581 (2009).}
\bibitem{Avgoustidis2010}{A. Avgoustidis et al., J. Cosmol. Astropart. Phys. 10, 024 (2010).}
\bibitem{Jaeckel2010}{J. Jaeckel and A. Ringwald, Annu. Rev. of Nuc. and Part. Sc.  60, 405 (2010).}
\bibitem{Hees2014}{A. Hees, O. Minazzoli and J. Larena, Phys. Rev. D 90, 124064  (2014).}
\bibitem{Holanda2013}{R. F. L. Holanda, J. C. Carvalho and J. S. Alcaniz, J. Cosmol. Astropart. Phys. 04, 027 (2013).}
\bibitem{DEBERNARDIS2006}{F. De Bernardis, E. Giusarma and A. Melchiorri, Int. J. Mod. Phys. D 15, 759 (2006).}
\bibitem{Uzan2004}{J. P. Uzan, N. Aghanim and Y. Mellier, Phys. Rev. D. 70, 083533 (2004).}
\bibitem{Holanda2011}{R. F. L. Holanda, J. A. S. Lima and M. B. Ribeiro Astron. \& Astrophys.  528, L14 (2011).}
\bibitem{Piazza2016}{F. Piazza and T. Schücker, Gen. Rel. and Grav. 48 (2016).}
\bibitem{Holanda2010}{R. F. L. Holanda, J. A. S. Lima and M. B. Ribeiro, Astrophys. J. 722, L233 (2010).}
\bibitem{Lima2011}{J. A. S. Lima, J. V. Cunha and V. T. Zanchin, Astrophys. J. 742, L26 (2011).}
\bibitem{Li2011}{Z. Li, P. Wu and H. Yu, Astrophys. J. 729, L14 (2011).}
\bibitem{gon}{R.S. Gonçalves, R. F. L. Holanda and J. S. Alcaniz, Mon. Not. Roy. Astron. Soc. 420, L43 (2012).}
\bibitem{Meng2012}{X.-L. Meng, T.-J Zhang, H. Zhan and X. Wang, Astrophys. J. 745, L98 (2012).}
\bibitem{Holanda2012}{R. F. L. Holanda, R. Gonçalves and J. Alcaniz, J. Cosmol. Astropart. Phys. 06, 022 (2012).}
\bibitem{Yang2013}{X. Yang, H.-R. Yu, Z.-S. Zhang and T.-J. Zhang, Astrophys. J. 777, L24 (2013).}
\bibitem{Liang2013}{N. Liang et al., Mon. Not. Roy. Astron Soc. 436, 1017 (2013).}
\bibitem{Shafieloo2013}{A. Shafieloo et al., J. Cosmol. Astropart. Phys. 04, 042 (2013).}
\bibitem{Zhang:2014eux}{Y. Zhang, (2014), arXiv:1408.3897 [astro-ph.CO].}
\bibitem{SantosdaCosta2015}{S. S. da Costa, V. C. Busti and R. F. L. Holanda, J. Cosmol. Astropart. Phys. 2015, 061 (2015).}
\bibitem{Jhingan2014}{S. Jhingan, D. Jain and R. Nair, Journal of Physics: Conference Series 484, 012035 (2014).}
\bibitem{Chen2015}{Z. Chen, B. Zhou and X. Fu, Int. J. of Theor. Phys.  55, 1229 (2015).}
\bibitem{Holanda2016}{R. F. L. Holanda and K. Barros, Phys. Rev. D 94, 023524 (2016).}
\bibitem{Rana2016}{A. Rana, D. Jain, S. Mahajan and A. Mukherjee, J. Cosmol. Astropart. Phys. 07, 026 (2016).}
\bibitem{Liao2016}{K. Liao et al., Astrophys. J. 822, 74 (2016).}
\bibitem{Holanda2016b} R. F. L. Holanda, V. C. Busti and J. S. Alcaniz, J. Cosmol. Astropart. Phys. 02, 054 (2016).
\bibitem{Holanda2017}{R. Holanda, S. Pereira and S. da Costa, Phys. Rev. D 95, 084006  (2017).}
\bibitem{Rana2017}{A. Rana, D. Jain, S. Mahajan, A. Mukherjee and R. F. L. Holanda, J. Cosmol. Astropart. Phys. 07, 010 (2017).}
\bibitem{Lin2018}{H.-N. Lin, M.-H. Li and X. Li, Mon.Not. Roy. Astron. Soc. 480, 3117 (2018).}
\bibitem{Fu2019}{X. Fu, L. Zhou and J. Chen, Phys. Rev. D 99, 083523  (2019).}
\bibitem{Ruan2018}{C.-Z. Ruan, F. Melia and T.-J. Zhang,  Astrophys. J. 866, 31 (2018).}
\bibitem{Holanda2019}{R. F. L. Holanda, L. Colaço, S. Pereira and R. Silva, J. Cosmol. Astropart. Phys. 06, 008 (2019).}
\bibitem{Chen2020}{J. Chen, Commun. in Theor. Phys. 72, 045401 (2020).}
\bibitem{Zheng2020}{X. Zheng et al.,  Astrophys. J. 892, 103 (2020).}
\bibitem{Kumar:2020ole}{D. Kumar, D. Jain, S. Mahajan, A. Mukherjee and N. Rani, Phys. Rev. D103, 063511 
(2021). }
\bibitem{Hu2018}{J. Hu and F.Y. Wang,  Mon.Not. Roy. Astron. Soc. 477 , 5064, (2018).}
\bibitem{HolandaWilliam}{W. J. C. da Silva, R. F. L. Holanda and R. Silva, Phys. Rev. D 102, 063513 (2020).}
\bibitem{Leaf2018lfu}{K. Leaf and F. Melia, Mon. Not.  Roy. Astron. Soc. 478, 4 (2018) }
\bibitem{Mantz2014}{A. B. Mantz et al., Mon. Not.  Roy. Astron. Soc. 440, 2077
(2014).}
\bibitem{Planelles2013}{S. Planelles et al., Mon.Not. Roy. Astron. Soc. 431, 1487 (2013).}
\bibitem{app} {D. E. Applegate  et al., Mon.Not. Roy. Astron. Soc. 457, 1522 (2016).
\bibitem{Cao2015} S. Cao et al., Astrophys. J. 806, 185 (2015).}
\bibitem{lentes}{P. Schneider, J. Ehlers and E. E. E. Falco, Springer-Verlag Berlin Heidelberg New york. Also Astronomy and Astrophysics Library 2019.}
\bibitem{Shu2017}{Y. Shu et al., Astrophys. J. 851, 1 (2017).}
\bibitem{Refsdal}S. Refsdal,  Mon.Not. Roy. Astron. Soc., 128, 307 (1964).
\bibitem{Treu}T. Treu and P. J. Marshall,  Astron. Astrophys. Rev., 24, 11 (2016) [arXiv:1605.05333].
\bibitem{Wong}K. C. Wong et al.,  Mon.Not. Roy. Astron. Soc. 498, 1420 (2020) [arXiv:1907.04869].
\bibitem{Liao}K. Liao et al., Astrophys. J. L. 886,  L23 (2019)
[arXiv:1908.04967].
\bibitem{Liao2}K. Liao et al., Astrophys. J. L. 895 L29 (2020) [arXiv:2002.10605].
\bibitem{kocha}{ C. S. Kochanek and P. L. Schechter. Published by Cambridge. University Press, as part of the Carnegie Observatories Astrophysics Series. Edited by W. L. Freedman, 117 (2004)  [arXiv:astro-ph/0306040].}
\bibitem{Holanda2017b}{R. F. L. Holanda, V. C. Busti, F. S. Lima and J. S. Alcaniz, J. Cosmol. Astropart. Phys. 09, 039 (2017) .}
\bibitem{rana}{A. Rana, D. Jain, S. Mahajan and A. Mukherjee, J. Cosmol. Astropart. Phys. 028, 03 (2017) }
\bibitem{gott}{J.R., Gott, M.-G. Park and H. M Lee, Astrophys. J. 338 , 8 ( 1989).}
\bibitem{jzqi}{J.-Z. Qi et al., Mon.Not. Roy. Astron. Soc. 483, 1 (2019) .}
\bibitem{fuku}{M. Fukugita, T. Futamase, M. Kasai and E. L. Turner, Astrophys. J. 393 1  (1992). }
\bibitem{luz}{S. Cao et al., Astrophys. J. 867, 50 (2018).}
\bibitem{bartel}{ M. Bartelmann and P. Schneider, Phys. Rep. 340, 291 (2001)}
\bibitem{Allen2008}{S. W. Allen et al.,  Mon. Not. Roy. Astron. Soc. 383, 879 (2008).}
\bibitem{Holsclaw}T. Holsclaw et al, Phys. Rev. D 82, 103502 (2010) [arXiv:1009.5443].
\bibitem{Shafieloo}A. Shafieloo, A. Kim and E. Linder, Phys. Rev. D 85, 123530 (2012) [arXiv:1204.2272].
\bibitem{seikel} M. Seikel, C. Clarkson and M. Smith,  J. Cosmol. Astropart. Phys.  06,  036 (2012).
\bibitem{elilensing}{ A. Sonnenfeld et al., Astrophys. J. 777, 98 (2013).}
\bibitem{grillo}{ C. Grillo, M. Lombardi and  G. Bertin,  Astron. \& Astrophys. 477, 397 (2008).}
\bibitem{holandafelipe} {R. F. L. Holanda et al., J. Cosmol. Astropart. Phys. 12,  016 (2017).}
\bibitem{holandaso}{R. F. L. Holanda, Astropart. Phys. 99, 1, (2018).}



\bibitem{melia} K. Leaf \& F. Melia2,  Mon. Not. R. Astron. Soc. 478, 5104 (2018)

\bibitem{wang} Z. L. Tu, J. Hu, \& F. Y. Wang1, Mon. Not.  Roy. Astron. Soc. 484, 4337 (2019).

\bibitem{amante} M. H. Amante et al., Mon. Not.  Roy. Astron. Soc. 498, 6013 (2020).

\bibitem{lizardo} A. Lizardo et al., Mon. Not.  Roy. Astron. Soc. 507, 5720 (2021).

\bibitem{cao} S. Cao et. al., The Astrophysical Journal, 806 185(2015).

\bibitem{wall} {J. V. Wall, \textit{Practical Statistics for Astronomers - I. Definitions, the Normal Distribution, Detection of Signal}, Quarterly Journal of the Royal Astronomical Society, 20, 138(1979).}

\bibitem{licia}{L. Verde, \textit{A practical guide to Basic Statistical Techniques for Data Analysis in Cosmology}, e-Print: 0712.3028 [astro-ph] (2007).}

\bibitem{sivia}{D.S.Sivia \& J.Skilling , \textit{Data Analysis: A Bayesian Tutorial}, Oxford University Press}




\bibitem{Caob}{S. Cao et al., Mon. Not. Roy. Astron. Soc. 457, 281 (2016).}
\bibitem{Schwab}{J. Schwab, A. S. Bolton, S. A. Rappaport,  Astrophys. J. 708, 750 (2010).}
\bibitem{Avgoustidis2012}{A. Avgoustidis et al., J. Cosmol. Astropart. Phys.  10, 024 (2010).}
\bibitem{Nair2012}{R. Nair, S. Jhingan and D. Jain, J. Cosmol. Astropart. Phys. 1105, 023 (2011) .}
\bibitem{Puxon2015}{W. Puxun, L. Zhengxiang, L. Xiaoliang and Y. Hongwei, Phys. Rev. D, 92, 023520 (2015).}
\bibitem{Fu2017}{X. Fu and P. Li,   
Inter. J. of Mod. Phys. D 26, 1750097 (2017).}
\bibitem{Li2018}{X. Li and H.-N. Lin, Mon. Not. Roy. Astron. Soc. 474, 1 (2017).}







\end{thebibliography}
\end{document}